\documentclass{appolb}

\usepackage{graphicx}
\usepackage{subfigure}
\usepackage{cite}
\usepackage{epsfig}
\usepackage{amsmath,amssymb}

\def\DeltaA{{\mathcal A}}
\newcommand{\ignore}[1]{}

\begin{document}

\title{Quark properties from the Hadron Resonance Gas\thanks{Talk by
    ERA at {\em Excited QCD 2015}, 8-14 March 2015 Tatranska Lomnica,
    Slovakia.}  \thanks{Report MPP-2015-100. Supported by Junta de
    Andaluc{\'\i}a grant FQM225, and the European Union 
%under a Marie Curie Intra-European Fellowship 
(FP7-PEOPLE-2013-IEF), project PIEF-GA-2013-623006.}
} 
\author{ E. Ruiz Arriola, L.L. Salcedo
  \address{Departamento de F\'{\i}sica At\'omica, Molecular y Nuclear \\
     and Instituto Carlos I de F\'{\i}sica Te\'orica y
    Computacional, \\ Universidad de Granada, E-18071 Granada, Spain}
  \and E. Meg\'{\i}as\address{Max-Planck-Institut f\"ur Physik (Werner-Heisenberg-Institut), F\"ohringer Ring 6, D-80805 Munich, Germany}
}
\maketitle
\begin{abstract}
We show how the quark free energy can be determined from a string and
the Hadron Resonance Gas model with one heavy quark below the
de-confinement phase transition. We discuss the interesting problem of
identification of degrees of freedom at increasing temperatures, as
well as the relevance of string breaking and avoided crossings.
\end{abstract}

 
\bigskip
 
\section{Introduction}

What is the maximum temperature where the bulk of QCD thermodynamics
can be described in hadronic terms, with no explicit reference to the
underlying quarks and gluons ? In this contribution we analyze to what
extent can one describe the QCD thermodynamics from the hadronic
spectrum. We thus hope to identify the nature of non-hadronic
precursors of the crossover to the quark-gluon plasma in terms of low
temperature partonic expansions~\cite{Megias:2012kb,Megias:2013xaa}
(see ~\cite{Arriola:2014bfa} for a review). The relation between
spectrum and partition function provides the thermodynamic free energy $F$ 
\begin{eqnarray}
Z (T)= \sum_n e^{-E_n/T} \equiv e^{-F/T} \, . 
\end{eqnarray}
The example of a Hydrogen gas in QED illustrates the situation and
motivates the discussion regarding the completeness of states which
ultimately are interacting protons and electrons. At low temperatures,
states are molecular, i.e. $H_2$ states which behave as compact and
structureless. When we heat up the system a rovibrational spectrum of
states $H_2^*, H_2^{**}$ emerges and eventually the molecule is
disociated into atoms $H_2 \leftrightarrow 2 H$. At higher
temperatures, atoms are excited $H^*,H^{**}, \dots$ and eventually
ionized into a plasma of $p$ and $e^-$ in the continuum where the $p$
and $e^-$ constituents becomes manifest. Clearly, neutral states
(atomic or molecular) are not effectively complete at all temperatures
and residual interactions corresponding to background non-resonant
scattering become relevant~\cite{rajaraman1979elementarity}.

\section{QCD vs the Hadron Resonance Gas}

\begin{figure}[t]
\subfigure{\includegraphics[angle=0,width=0.48\textwidth]{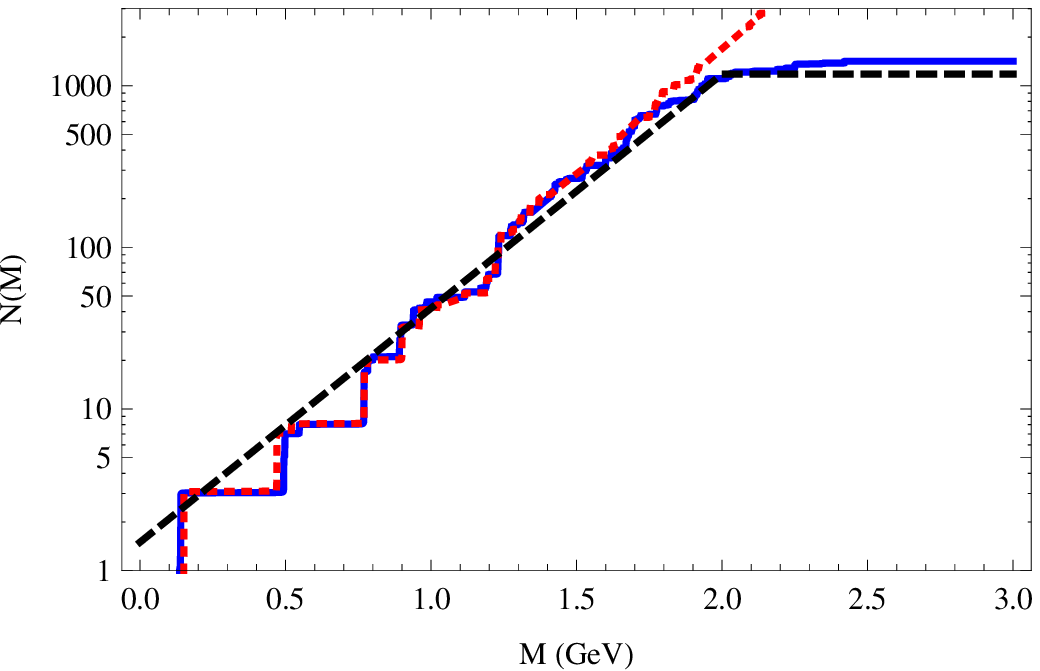}} 
\hfill
\subfigure{\includegraphics[angle=0,width=0.48\textwidth]{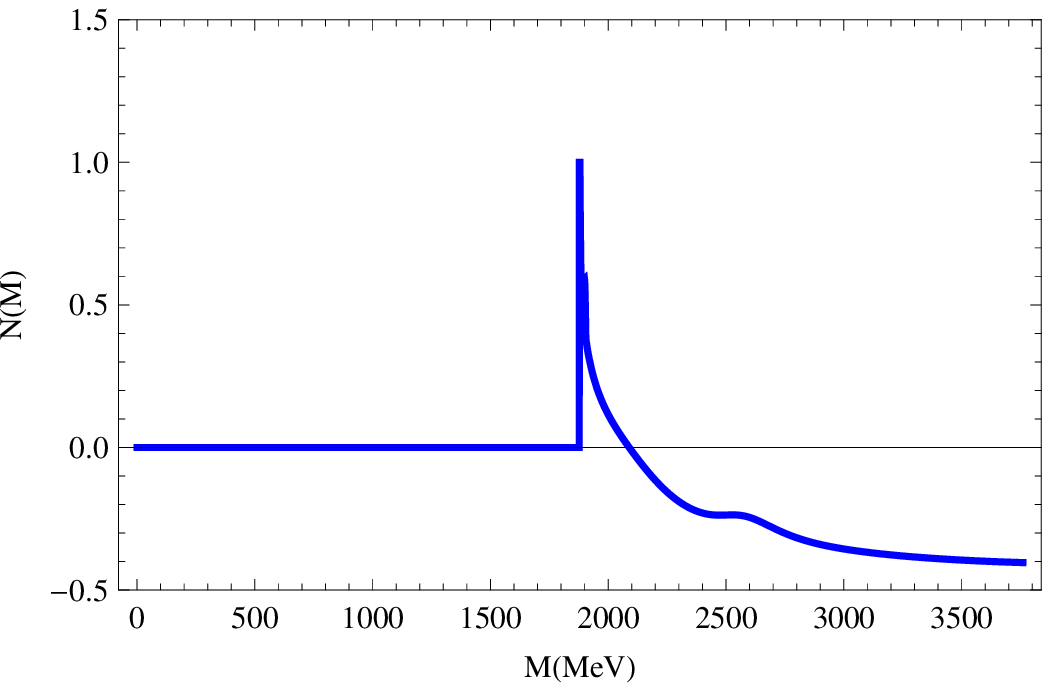}} %\hfill
%\subfigure{\includegraphics[angle=0,width=0.49\textwidth]{figtraceanomaly.eps}}
%\epsfig{figure=figcumulative.eps,height=4cm,width=6cm,angle=0}
%\epsfig{figure=figtraceanomaly.eps,height=4cm,width=6cm,angle=0} 
%\epsfig{figure=figCoVolumeIsgur1.eps,height=4cm,width=6cm,angle=0}
\caption{Left: Total cumulative number $N(M)=\sum_{n} \theta(M-M_n)$
  for PDG (full), RQM (dotted) and a fit (dashed), Eq.~(\ref{eq:chi2})
  with $N(M)=A [e^{M/T_H} \theta(M_{\rm th}-M) + e^{M_{\rm th}/T_H}
    \theta(M-M_{\rm th}) ]$. For $M_{\rm th}= 2 {\rm GeV}$ gives
  $A=1.5$ and $T_H=300 {\rm MeV}$ with $\chi^2/\nu=0.94 $ and $T_{\rm
    max}=185 {\rm MeV}$. Right: The single contribution of the deuteron
  $^3S_1$-channel to $N(M)$ as a function of
  the invariant NN mass.}
\label{fig:deut}
\end{figure}

In QCD on the lattice $F$ is determined by integrating the trace
anomaly,
\begin{eqnarray}
\DeltaA (T) \equiv \frac{\epsilon - 3 P}{T^4} = T \frac{\partial}{\partial T}
\left( \frac{P}{T^4} \right) \,,  
\label{eq:em3p}
\end{eqnarray}
with respect to the temperature and {\it assuming} a reference value,  
ideally $P(T_0)= P_{\pi}(T_0)$ with $T_0 \ll m_\pi$. Once this
condition is imposed we expect that at sufficiently low
temperatures all the observables are described in terms of hadronic
degrees of freedom; the quark-gluon underlying constituents are
disclosed at very high temperatures. This quark-hadron duality {\it at
  low temperature} has been confirmed by the most recent and accurate
lattice calculations at finite
temperature~\cite{Borsanyi:2013bia,Bazavov:2014pvz} are confronted
with the Hadron Resonance Gas (HRG), a rather simple system of
non-interacting point-like and structureless hadrons
\begin{eqnarray}
\DeltaA_{\rm HRG} (T) =
%\equiv \frac{\epsilon - 3 P}{T^4} =
\frac{1}{T^4}\int dN(M) \int \frac{d^3 p}{(2\pi)^3} \frac{E(p)- \vec p
  \cdot \nabla_p E_(p)}{e^{E(p)/T}+\eta} \, .
\label{eq:tran-HRG}
\end{eqnarray}
Here, the sum is over {\it all} hadronic states including spin-isospin
and anti-particle degeneracies, $\eta=\mp 1$ for mesons and baryons
respectively, $E(p)= \sqrt{p^2+M^2}$ is the energy and the cumulative
number of states (degeneracy included) $N(M)= \sum_n \theta (M-M_n)$ where $M_n$ are the
hadron masses. The spectrum can be taken either directly from the
PDG~\cite{Nakamura:2010zzi} or the relativized quark model
(RQM)~Refs.~\cite{Godfrey:1985xj,Capstick:1986bm}.
%The partition
%function is actually reconstructed from the trace anomaly with a
%suitable boundary condition at low temperature where pions dominate.
%In terms of ${\cal A}$ the agreement of both the PDG and the RQM is
%with lattice is very good. 
Indeed, requiring 
\begin{eqnarray}
\chi^2 = \sum_{i=1}^{N_{\rm dat}} \left[\frac{{\cal A}_{\rm Lat} (T_i) - {\cal
    A}_{\rm HRG} (T_i)}{\Delta {\cal A}_{\rm Lat} (T_i)} \right]^2 \sim N_{\rm dat}
 \pm \sqrt{2 N_{\rm dat}} \, , 
\label{eq:chi2}
\end{eqnarray}
we get $T \lesssim 175 {\rm MeV}$ for $N_{\rm dat}=10$ lattice
points~\cite{Borsanyi:2013bia,Bazavov:2014pvz} both for PDG and RQM,
showing that they can be used to saturate the hadronic spectrum for
light quarks provided $M \lesssim 2.1 {\rm GeV}$. This RQM saturation
will be assumed also to take place when heavy quarks are involved in
which case the PDG lacks many of the needed states. This of course
suggests to determine {\it directly} $N(M)$ from the ${\cal A}$ data,
as illustrated in Fig.~\ref{fig:deut} (left panel).

\section{Counting hadrons}

The partition function and its comparison with hadronic states through
the trace anomaly ${\cal A}$, displays explicitly the connection
between spectrum and thermodynamics yielding the statistically
significant identification of $N_{\rm QCD}$ with $N_{\rm HRG}$ using
PDG or RQM where only $\bar q q $ and $qqq$ states contribute. But,
which hadronic states count into the calculation of the thermodynamic
properties ?. As argued long ago~\cite{Dashen:1974ns}, counting
hadronic states implicitly averages over some scale, and so states
such as the deuteron (made of $6q$) generate fluctuations in a smaller
nuclear scale. The cumulative number in a given channel $\alpha$ with
bound states $M_{n,\alpha}$ below the threshold $M_{\rm th}$ is in
general (see e.g. \cite{rajaraman1979elementarity})
\begin{eqnarray}
N_\alpha (M)= \sum_n \theta(M-M_{n,\alpha}) + \frac1{\pi} \left[\delta_\alpha(M)-\delta_\alpha (M_{\rm th})\right] \, , 
\end{eqnarray}
which becomes $N(\infty)=n_B + [\delta(\infty)-\delta(M_{\rm
    th})]/\pi=0$ due to Levinson's theorem. This fluctuation is shown
in Fig.~\ref{fig:deut} for the NN channel where $M_{\rm th}= 2M_N$,
and the deuteron mass is $M_d= 2 M_N -B_d $ ($B_d=2.2 {\rm
  MeV}$). Thus, the deuteron doesn't count. This is a general feature
of weakly bound states, which for the copious new X,Y,Z states, might
affect the thermodynamics if the HRG was blindly identified with the
PDG (with {\it all} X,Y,Z's).

\section{Quark potential, string breaking and avoided crossing}

\begin{figure}[t]
\subfigure{\includegraphics[angle=0,width=0.48\textwidth]{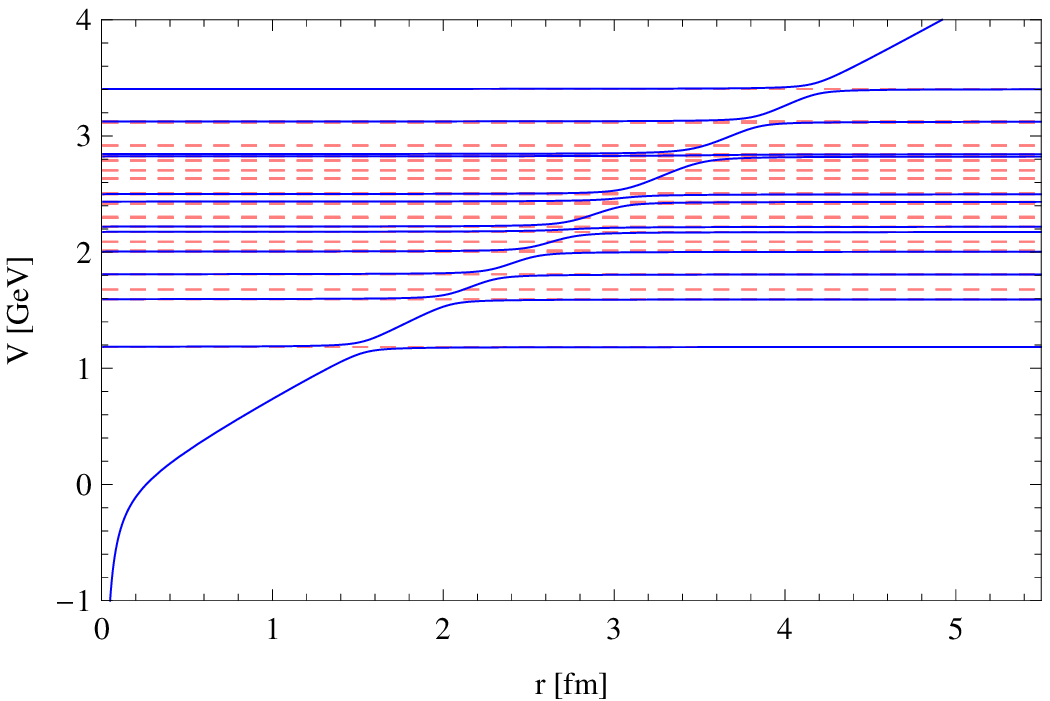}} 
\hfill
\subfigure{\includegraphics[angle=0,width=0.48\textwidth]{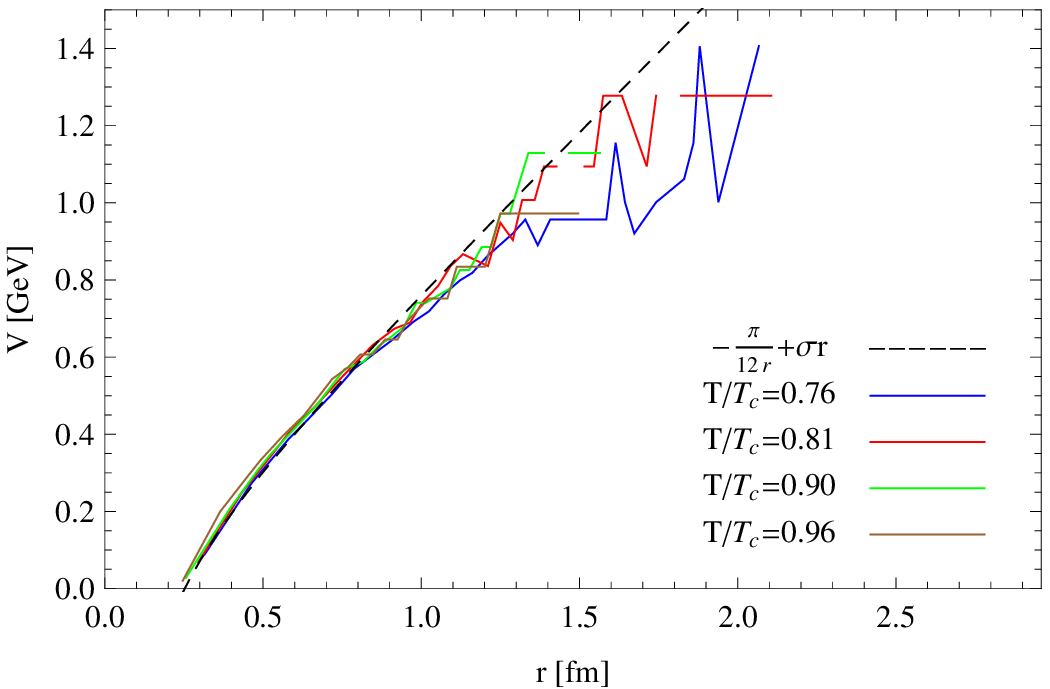}} 
\caption{\footnotesize Left:Potential as a function of distance for the
  model including 
heavy-light mesons with a
  charm quark~\cite{Godfrey:1985xj}, up to $\Delta=1.8\,$ GeV. We
  include only a $g= 50\,$MeV mixing between the fundamental and
  the excited states.  Right: Heavy
  $\bar{Q}Q$ potential as a function of distance extracted from the
  lattice data of Ref.~\cite{Kaczmarek:2005ui} for temperatures
  $T/T_c= 0.76,\, 0.81,\, 0.90$ and $0.96$ using
  Eq.~(\ref{eq:Fave2}). The dashed (red) line is the result for the
  $\bar{Q}Q$ Cornell potential of Eq.~(\ref{eq:Cornell}) with
  $\sigma=(0.4 \, \textrm{GeV})^2$.}
\label{fig:Isgur3}
\end{figure}

The linearly growing static energy of two $\bar Q $ and $Q $ sources
in the fundamental representation of the SU($N_c$)-group placed at a
distance $r$ is often identified with confinement with a string
tension $\sigma = (0.42 {\rm GeV})^ 2$. Including the short
distance (perturbative) Coulomb-like behaviour yields the
venerable Cornell potential for the ground state
\begin{eqnarray}
V_{Q \bar Q} (r) &=& \sigma \, r - \frac{4 \alpha_s}{3r} + \cdots \,, 
\label{eq:Cornell}
\end{eqnarray}
Actually, the mass of a $\bar Q Q$ state, becomes unstable at a
critical distance $r_c$ when a light $\bar q q$ pair is created from
the vacuum and the heavy-light meson-antimeson $\bar B B \equiv (\bar
Q q)(Q \bar q)$ channel opens
\begin{eqnarray}
M_{\bar Q Q} (r_c) = V_{Q \bar Q} (r_c) +m_{\bar Q} + m_{Q} =
M_{\bar B} + M_{B} \, , 
\end{eqnarray}
where the weak van der Waals like meson exchange interaction in the
$\bar B B$ sector is neglected~\footnote{A rough estimate of $r_c$
  proceeds as follows.  Due to spontaneous chiral symmetry breaking
  the quarks acquire an additive {\it dynamical} mass $M_0$
  contribution to the total mass $M_i= M_0 + m_i$ where $m_i$ is the
  {\it current} mass. For light $\bar q q$-mesons $M_\rho = 2 M_0 +
  2m_q$ whereas for light-heavy ($\bar q Q$) mesons $M_{B} = 2 M_0 +
  m_Q + m_{\bar q}$ and thus $ V_{Q \bar Q} (r_c) = 4 M_0 + 2 m_q $
  which yields $r_c \sim 4 M_0 /\sigma \sim 1.2-1.4 {\rm fm}$ in fair
  agreement with the lattice estimate~\cite{Bali:2005fu}.}. We have a
diabatic crossing structure which turns into an adiabatic avoided
crossing when the transition strength $V_{\bar Q Q \to \bar B B}$ is
included. In general, one has excited meson states, $V^{(n,m)}_{\bar Q
  q , \bar q Q}(r) = \Delta^{(n)}_{q\bar{Q}}+ \Delta^{(m)}_{\bar{q}Q}$
where $\Delta^{(n)}_{\bar{q}Q}= M_B^{(n)} - m_Q$ and in
Fig.~\ref{fig:Isgur3} we show the resulting adiabatic spectrum based
on the RQM~\cite{Godfrey:1985xj} when a mixing strength of $50 {\rm
  MeV}$ is implemented.

%\begin{eqnarray}
%E_0(r) &=& \sigma r \theta (r_c-r ) + 4 M_0 \theta (r-r_c) \,,
%\\ E_0^*(r) &=& 4 M_0 \theta (r_c-r) + \sigma r \theta (r-r_c) \,,
%\label{eq:string-breaking}
%\end{eqnarray}
%
%\begin{eqnarray}
%V^{(0,0)}_{\bar Q Q} (r) &=& -\frac{4\alpha}{3 r}+ \sigma r \,, \nonumber \\ 
%V^{(n,m)}_{\bar Q q , \bar q Q}(r) &=& \Delta^{(n)}_{q\bar{Q}}+ \Delta^{(m)}_{\%bar{q}Q} \,, \nonumber 
%\label{eq:string-breaking-bis}
%\end{eqnarray}
%where $ \Delta^{(n)}_{q\bar{Q}} = M^{(n)}_{q\bar{Q}} -
%m_{\bar{Q}}$.  

\section{Heavy $\bar QQ$ free energy}

\begin{figure}[ht]
\begin{center}
\epsfig{figure=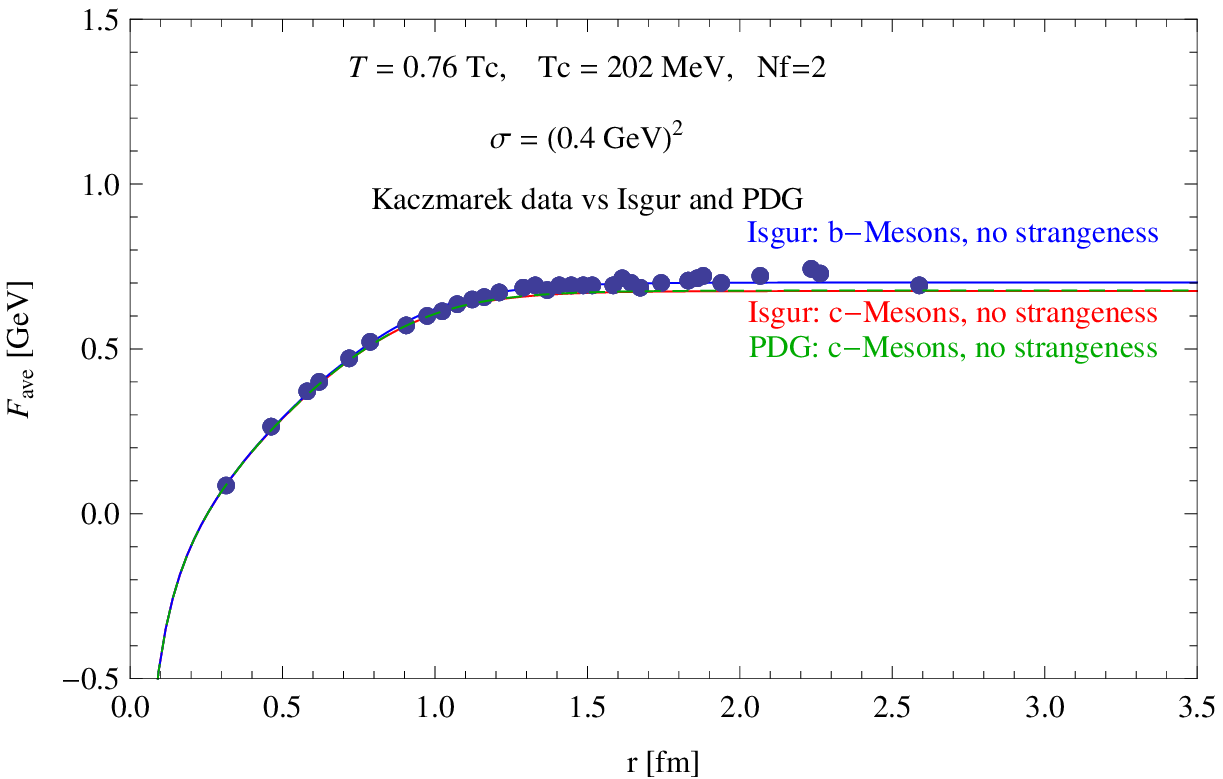,height=50mm,width=60mm}
\epsfig{figure=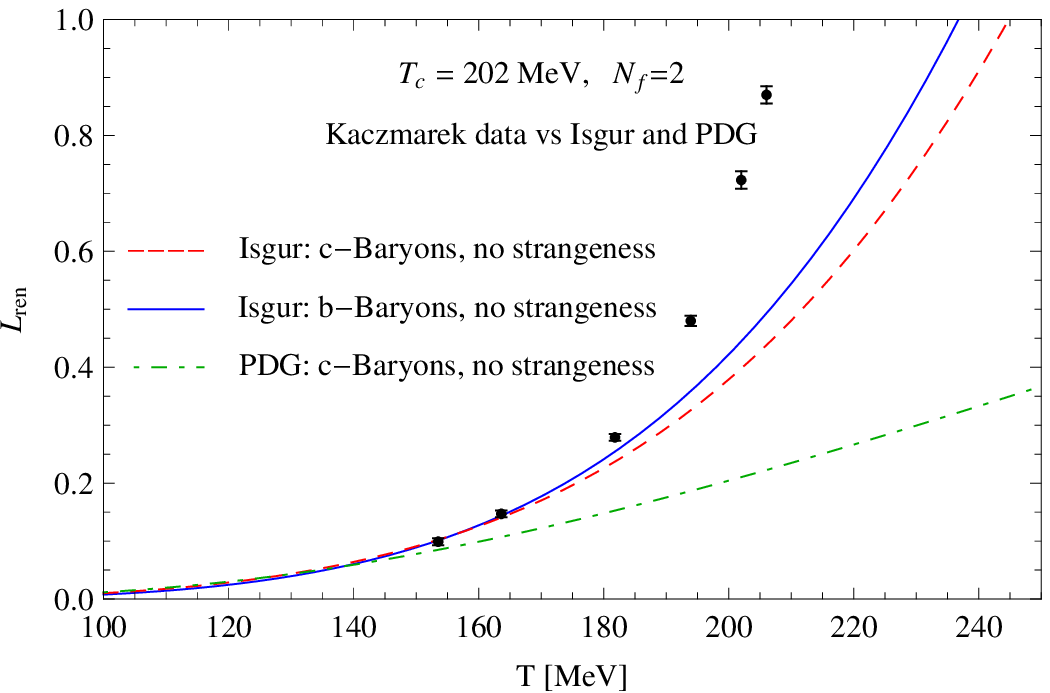,height=50mm,width=60mm}
\end{center}
\caption{\small Left: Color averaged heavy $\bar{Q}Q$ free energy as a
  function of distance. We show as dots the lattice data for 
  $N_f=2$, $T=0.76T_c$ taken from Ref.~\cite{Kaczmarek:2005ui}, and as
  continuous lines the result by using Eq.~(\ref{eq:Fave2}) with the
  spectrum of heavy-light mesons with a charm quark (red) and bottom
  quark (blue), with no strangeness, obtained with the Isgur model of
  Ref.~\cite{Godfrey:1985xj}, as well the spectrum obtained from the
  PDG (green). Right: Renormalized Polyakov loop as a function of
  temperature. We show as dots the lattice data for $T=0.76T_c$ taken
  from Ref.~\cite{Kaczmarek:2005ui}, and as continuous lines the
  result by using Eq.~(\ref{eq:Lren}).}
\label{fig:FLIsgurMesons}
\end{figure}

Much of our understanding of the QCD dynamics at finite temperature is
linked to the Helmholtz free energy defined as the maximum work the
system can exchange at fixed temperature between two heavy $\bar QQ$
at separation $r$.  In the confined phase, the correlator between
Polyakov loops becomes~\cite{Arriola:2014bfa}
\begin{equation}
e^{-F_{\rm ave}(r,T)/T} = \langle {\rm Tr}_F \Omega (\vec r) {\rm Tr}_F \Omega(0)^\dagger \rangle = \sum_{n,m}  e^{-V_{\bar{Q}Q}^{(n,m)}(r)/T}  \,, \label{eq:Fave}
\end{equation}
where, before mixing, one has the crossing among the energy levels up
to $\bar{q}q$ pair creation. Neglecting the avoided crossing,  
Eq.~(\ref{eq:Fave}) yields
\begin{equation}
e^{-F_{\rm ave}(r,T)/T}  = e^{-V_{\bar Q Q}(r)/T} + \left(\frac{1}{2}\sum_n e^{-\Delta_n /T} \right)^2 \,, \label{eq:Fave2}
\end{equation}
where $\Delta_n = \Delta^{(n)}_{q\bar{Q}} =
\Delta^{(n)}_{\bar{q}Q}$. We show in Fig.~\ref{fig:FLIsgurMesons} the
heavy $\bar{Q}Q$ free energy obtained with Eq.~(\ref{eq:Fave2}) by
using the spectrum of heavy-light mesons with a charm (bottom) quark,
and no strangeness, obtained with the Isgur model of
Ref.~\cite{Godfrey:1985xj} up to $\Delta = 3.19\,$GeV. We have
considered $\sigma=(0.4 \,\textrm{GeV})^2$ and $\alpha = \pi/16$ (no
fit).
%Note that we are not performing any fit to the lattice data. 
In Fig.~\ref{fig:FLIsgurMesons} we also show the Polyakov loop computed
as
\begin{equation}
L(T) \equiv \lim_{r\to\infty}  e^{-F_{\rm ave}(r,T)/(2T)}  =  \frac{1}{2}\sum_n e^{-\Delta_n /T}  \,. \label{eq:Lren}
\end{equation}
The agreement with lattice data is quite good for $T < 0.8 T_c$.  Thus
the spectrum both for the PDG and the RQM saturates the sum rules at
these temperatures~\footnote{By multiplying $L(T)$ by a factor
  $e^{C/T}$ with $C=-40$MeV, we get good agreement with the lowest
  temperature lattice point. This kind of ambiguity comes from
  renormalization effects. Like in~\cite{Kaczmarek:2005ui} we assume
  $F_{\rm ave} (r,T) \sim V_{\bar Q Q} (r) $ for $r T \ll 1$.  The
  recent evaluation of Ref.~\cite{Borsanyi:2015yka},
  unlike~\cite{Kaczmarek:2005ui} is based on a different
  renormalization condition and $N_f=2+1$ for the free energy and the role of
  avoiding crossing will be analyzed elsewhere.} whereas the finite
temperature does not resolve avoided crossings making them effectively
diabatic passages.

\section{Conclusions}

While much of the effort has been directed towards a description of
the hadron-to-quark-gluon crossover, at present there is no
understanding on the mechanism, perhaps because confinement itself is
not really well understood. Our analysis of the free energy suggests
that lowest temperatures where purely hadronic states account for QCD
observables is rather large $T \sim 170-180 {\rm MeV}$ and close to
the critical temperature.

%\bibliography{Refs,refs2,excitedQCD}
%\bibliographystyle{h-elsevier}

\end{document}